\documentclass[12pt]{iopart}
\usepackage{graphicx}

\newcommand*{\mub}{\ensuremath{\mu_{B}}}

\newcommand*{\ktopi}{\ensuremath{(\mathrm{K}^{+}\!+\mathrm{K}^{-})/(\pi^{+}\!+\pi^{-})}}
\newcommand*{\ptopi}{\ensuremath{(\mathrm{p} + \bar{\mathrm{p}})/(\pi^{+}\!+\pi^{-})}}
\newcommand*{\ktop}{\ensuremath{(\mathrm{K}^{+}\!+\mathrm{K}^{-})/(\mathrm{p}+\bar{\mathrm{p}})}}
\newcommand*{\ktopplus}{\ensuremath{\mathrm{K}^{+}\!/\mathrm{p}}}
\newcommand*{\roots}{\ensuremath{\sqrt{s_{_{NN}}}}}
\newcommand*{\sdyn}{\ensuremath{\sigma_{\mathrm{dyn}}}}

\newcommand*{\gev}{\ensuremath{\mathrm{GeV}}}

\begin{document}

\title[Ratio fluctuations in Pb+Pb collisions at SPS energies]{New results on event-by-event ratio fluctuations in Pb+Pb collisions at CERN SPS energies}

\author{Tim~Schuster for the NA49 collaboration}

\address{Fachbereich Physik der Universit\"{a}t and FIAS, Frankfurt, Germany.}

\ead{Tim.Schuster@cern.ch}

\begin{abstract}
Event-by-event fluctuations of produced particle multiplicities are believed to be sensitive to a deconfinement phase transition and the critical point of strongly interacting matter. The NA49 collaboration has conducted a systematic study of various fluctuation observables.
In this contribution, recent results on hadron ratio fluctuations in central Pb+Pb collisions at $\roots = 6.3$ to $17.3~\gev$ are reported. These results are complemented by the centrality dependence of hadron ratio fluctuations at $\roots = 17.3~\gev$. A universal scaling was found, describing the energy and centrality dependence of the \ktopi\ and \ptopi\ ratio fluctuations purely by a change in the observed average hadron multiplicities. This scaling is broken for the fluctuations of the \ktop\ and \ktopplus\ ratios, possibly hinting at a change in the baryon number-strangeness correlation.
\end{abstract}


An event-by-event measurement of hadron ratios allows to characterize the hadro-chemical composition of the ``fireball'' of hot and dense nuclear matter created in individual heavy-ion collisions.
In the vicinity of a phase transition, where the underlying degrees of freedom change, distinct fluctuation patterns in these ratios are expected.
This is substantiated by lattice QCD calculations~\cite{Karsch:2003QGP}, that exhibit a swift change in quark number susceptibilities at the critical temperature for vanishing baryo-chemical potential. Recent results at finite \mub~\cite{Karsch:2007dt,Karsch:2007dp} even report diverging susceptibilities, that are possibly related to the critical point of strongly interacting matter.

An additional motivation to study hadron ratio fluctuations is the expected change of the baryon-strangeness correlation~\cite{Koch:2005vg} at the deconfinement phase transition. This might be reflected in the fluctuations of the K/p ratio, because kaons and protons are the major carriers of strangeness and baryon number.

NA49~\cite{Afanasev:1999iu} is a large acceptance fixed target hadron spectrometer at the CERN Super Proton Synchrotron (SPS).
The main detectors used in the present analysis are the Time Projection Chambers (TPCs), used for tracking and particle identification (PID) via specific energy loss. Their large acceptance makes event-by-event measurements possible, as a significant fraction of the produced hadrons can be observed and identified.

A good knowledge of this acceptance is crucial: While for inclusive measurements it is possible to extrapolate into unmeasured regions, this is not possible in the case of fluctuations. For the interpretation of fluctuation results it is thus important to compare to models within the same acceptance. For that reason tables describing the NA49 acceptance are available at~\cite{acc_tab_edms}, and all model comparisons reported here have been performed within this acceptance.

Particle identification with the NA49 TPCs is done in the relativistic rise region of the specific energy loss in the detector gas.
Event-wise hadron ratios are extracted with a maximum likelihood method. A detailed description of this method can be found in~\cite{Gazdzicki:1994vj,Afanasev:2000fu}.
The scaled width of a resulting hadron ratio (e.g.\ $K/\pi$) distribution, defined as $\sigma^2 = \frac{\mathrm{Var}\left(K/\pi\right)}{\langle K/\pi \rangle^2}$ now contains, besides the physics correlations to be studied, additional effects from the PID method and from finite number statistics. The latter two are reproduced using a reference sample by subjecting mixed events to the same PID method as the data.
The ``dynamical'' fluctuations, arising only from the physics correlations, can then be calculated as

\begin{equation}
\sigma_{\mathrm{dyn}}=\mathrm{sign} \left(\sigma^2_{\mathrm{data}} - \sigma^2_{\mathrm{mix}} \right) \sqrt{\left|\sigma^2_{\mathrm{data}} - \sigma^2_{\mathrm{mix}} \right|}.
\end{equation}

\begin{figure}
\includegraphics[width=5cm]{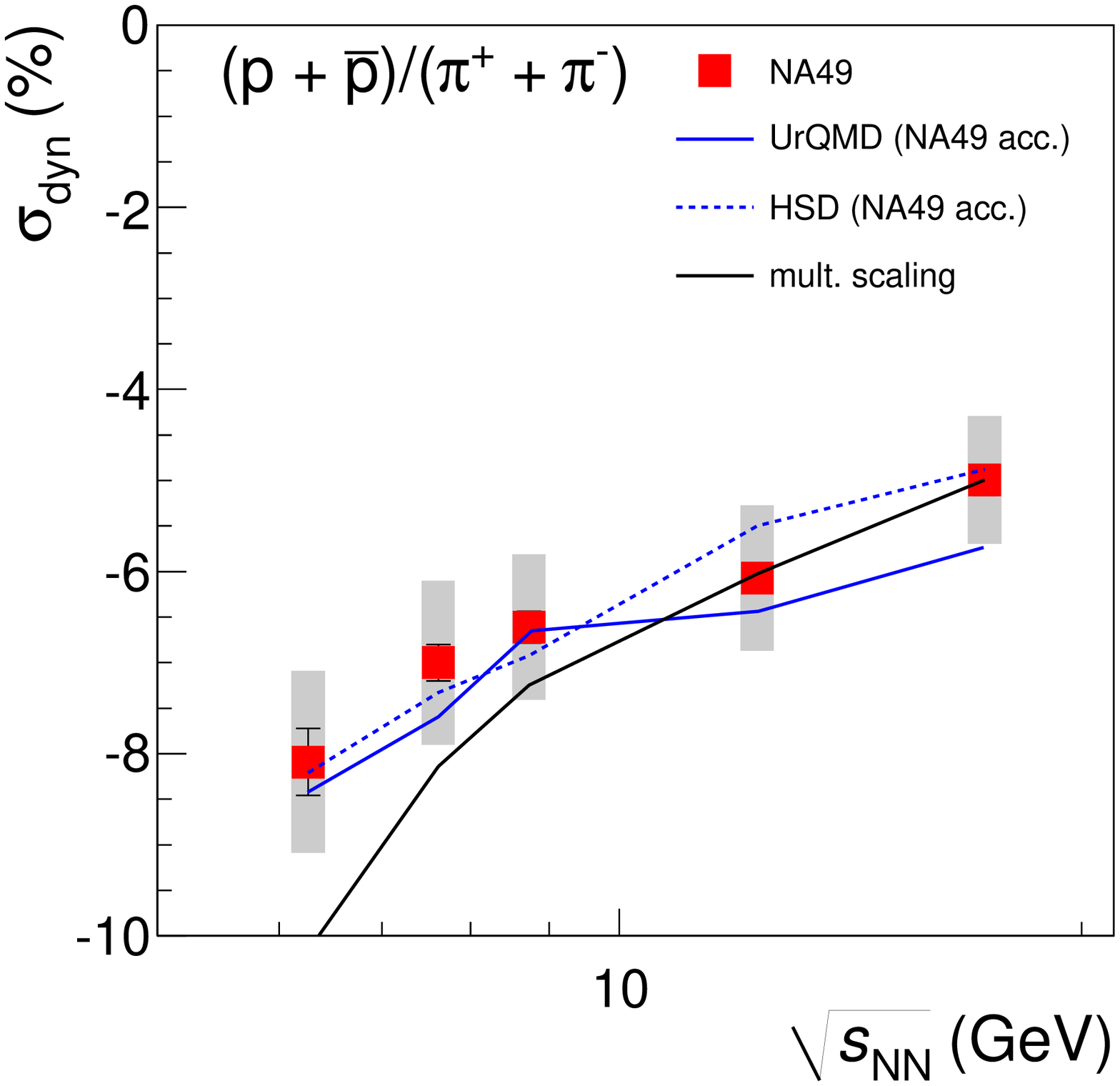}
\includegraphics[width=5cm]{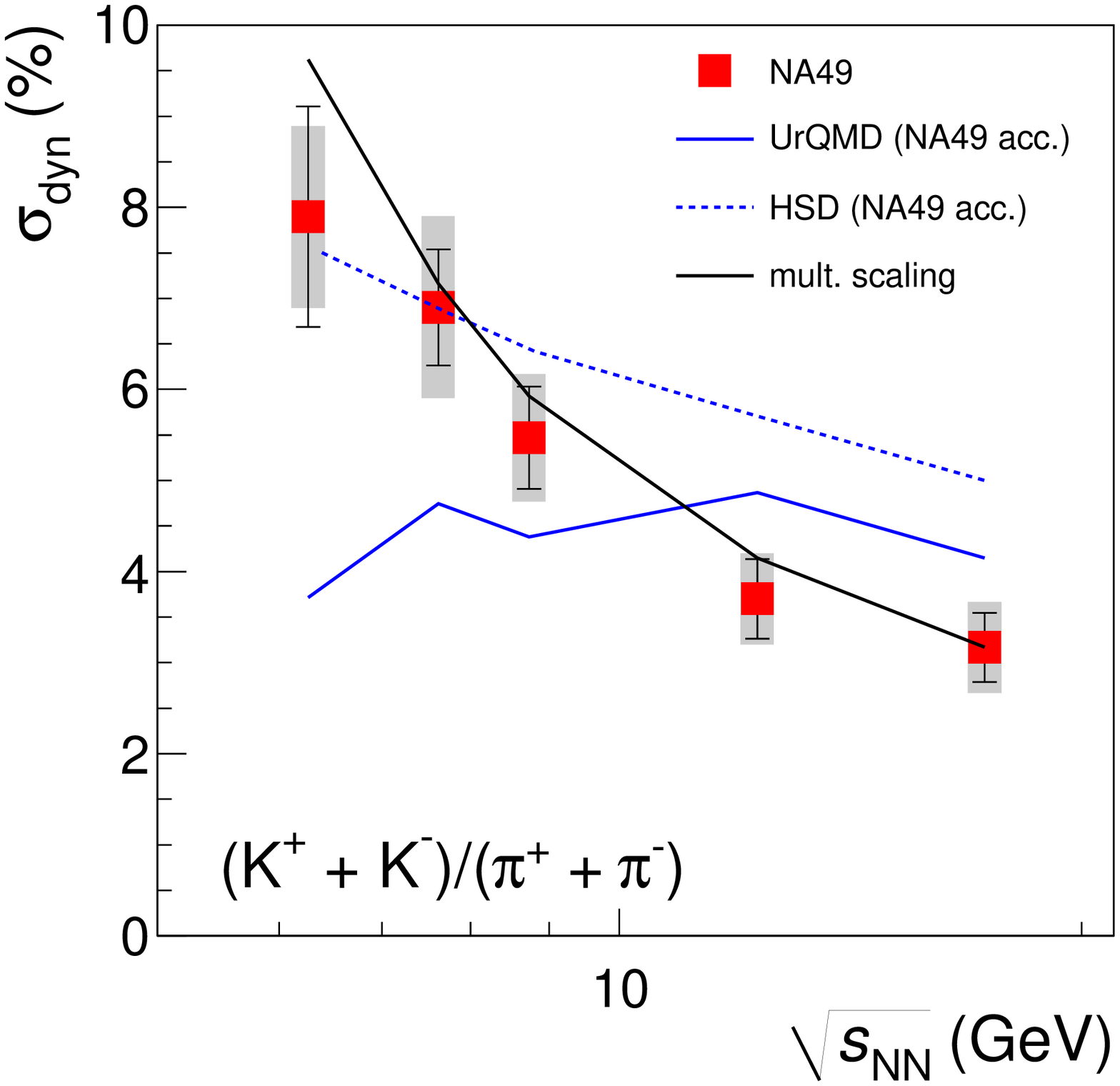}
\caption{Energy dependence of \sdyn\ for \ptopi\ and \ktopi\ in central Pb+Pb collisions, compared to results from the transport models UrQMD and HSD, as well as to the multiplicity scaling described in the text.}
\label{fig:Exc_PrPi_KPi}
\end{figure}

\sdyn\ of the \ptopi\ and \ktopi\ ratios was studied in central Pb+Pb collisions at five energies between 6.3 and 17.3 GeV~\cite{:2008ca}. The results are shown in figure~\ref{fig:Exc_PrPi_KPi}, with vertical bars indicating the statistical errors, while the grey boxes represent the systematic errors determined in extensive checks of the analysis.
For \ptopi\ negative values of \sdyn\ are observed, indicating a correlated production. The data is well described by the hadronic transport models UrQMD~\cite{Petersen:2008kb} and HSD~\cite{Konchakovski:2009at} with the production and strong decay of nucleon resonances as the dominating process correlating proton and pion numbers.
For \ktopi\ fluctuations positive values of \sdyn\ and a rise towards low energies is observed. The energy dependence of this anticorrelation is not reproduced by the hadronic models. UrQMD describes the high energy points, but shows only a weak change towards low energies. In HSD, the rise towards low energies can be observed, but the high energy points are over-predicted.

An alternative explanation of the energy dependence however can be derived from properties of the observable \sdyn\ itself. It can be separated~\cite{Koch:2009dg} into a correlation strength term and one which is purely dependent on multiplicities. If we assume an invariant correlation strength, as e.g.\ realized in a grand-canonical hadron resonance gas at fixed temperature and chemical potentials, only the multiplicity dependence remains.
The general expectation from~\cite{Koch:2009dg} is
\begin{equation}
\sigma_{\mathrm{dyn}} \propto \sqrt{\frac{1}{\langle A \rangle} + \frac{1}{\langle B \rangle}},
\label{eq:sc}
\end{equation}
for invariant correlation strength, where $\langle A \rangle$ and $\langle B \rangle$ are the average multiplicities of the two hadron species under consideration. This scaling, indicated by the black lines in figure~\ref{fig:Exc_PrPi_KPi} works surprisingly well in both cases, \ptopi\ and \ktopi.

\begin{figure}
\includegraphics[width=15cm]{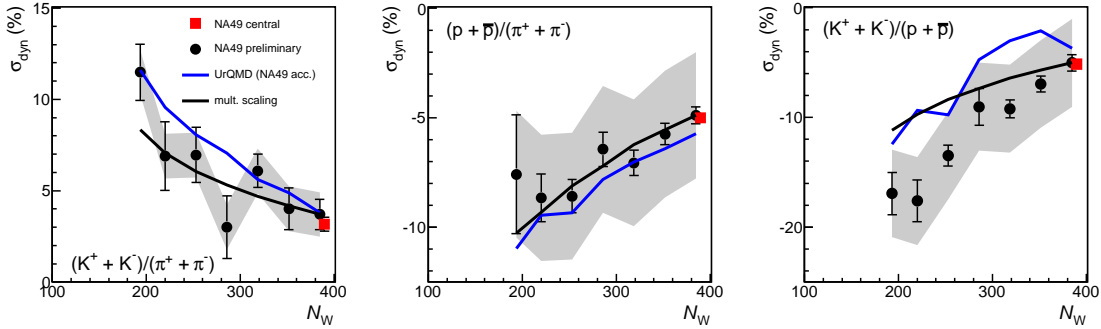}
\caption{Centrality dependence of \ktopi, \ptopi\ and \ktop\ fluctuations in Pb+Pb collisions at $\roots = 17.3$~GeV. The centrality is expressed in terms of $N_{\mathrm{W}}$, the number of ``wounded'' or participating nucleons.}
\label{fig:AllCent}
\end{figure}

Complementary to the energy scan, the centrality dependence of \ptopi, \ktopi\ and \ktop\ fluctuations in Pb+Pb collisions at the top SPS energy of $\roots = 17.3$~GeV was studied~\cite{NA49:Cent} and is presented in figure~\ref{fig:AllCent}.
The scaling properties of \sdyn\ found for the energy dependence above can thus be further tested by an analysis at fixed energy, only varying the system size.
This has been done before by STAR~\cite{:2009if} for \ktopi\ fluctuations at two higher energies, and an increase of \sdyn\ when going to peripheral collisions was observed.
All three ratios studied by NA49 show a similar increase in the magnitude of \sdyn\ when going to peripheral collisions.
The hadronic transport model UrQMD predicts a similar behaviour here, and the scaling suggested above shows the same dependence. This is compatible with the hypothesis that at constant energy the underlying correlations are not significantly changed by a variation of the system size.

\begin{figure}
\includegraphics[width=5cm]{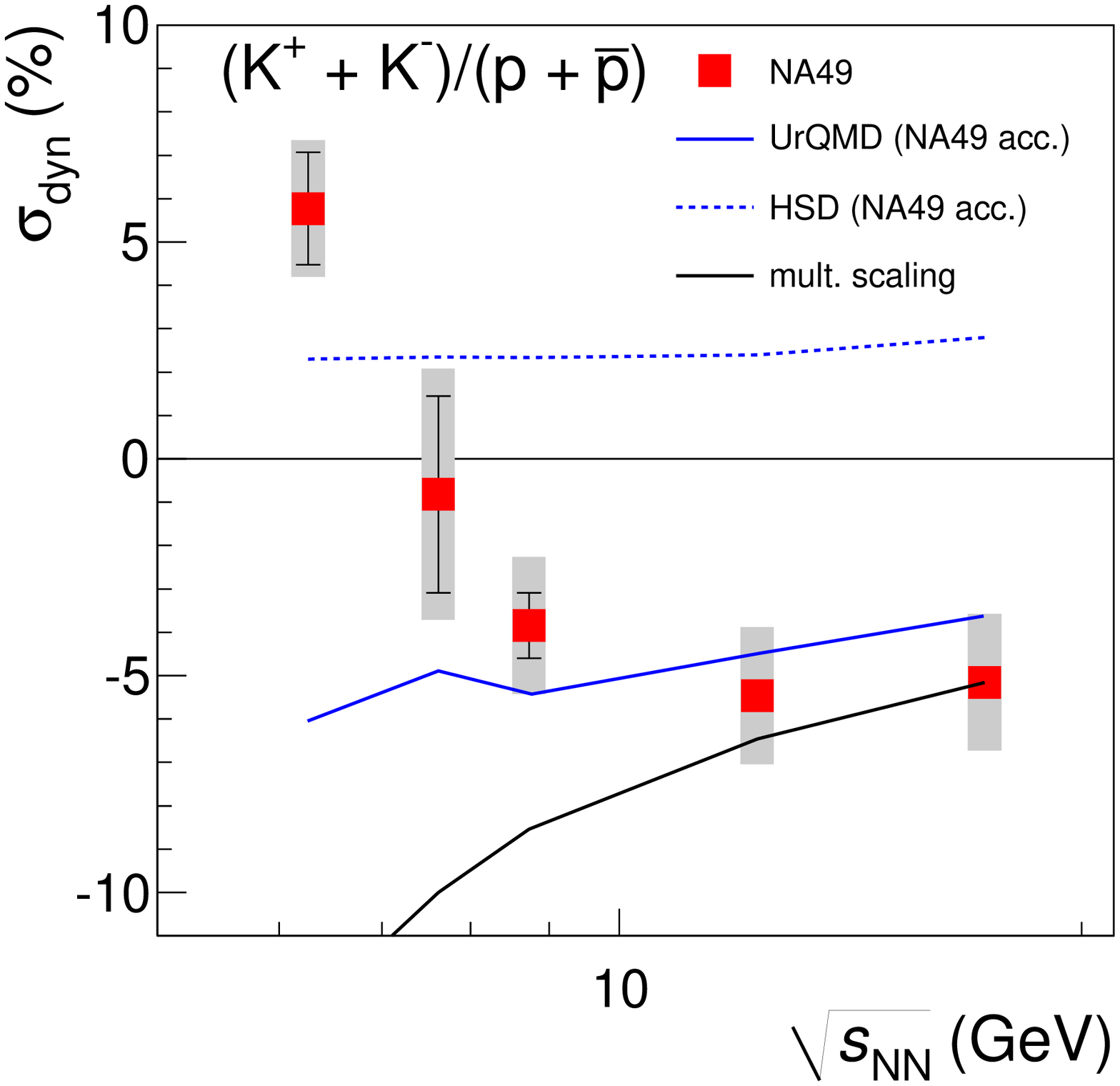}
\includegraphics[width=5cm]{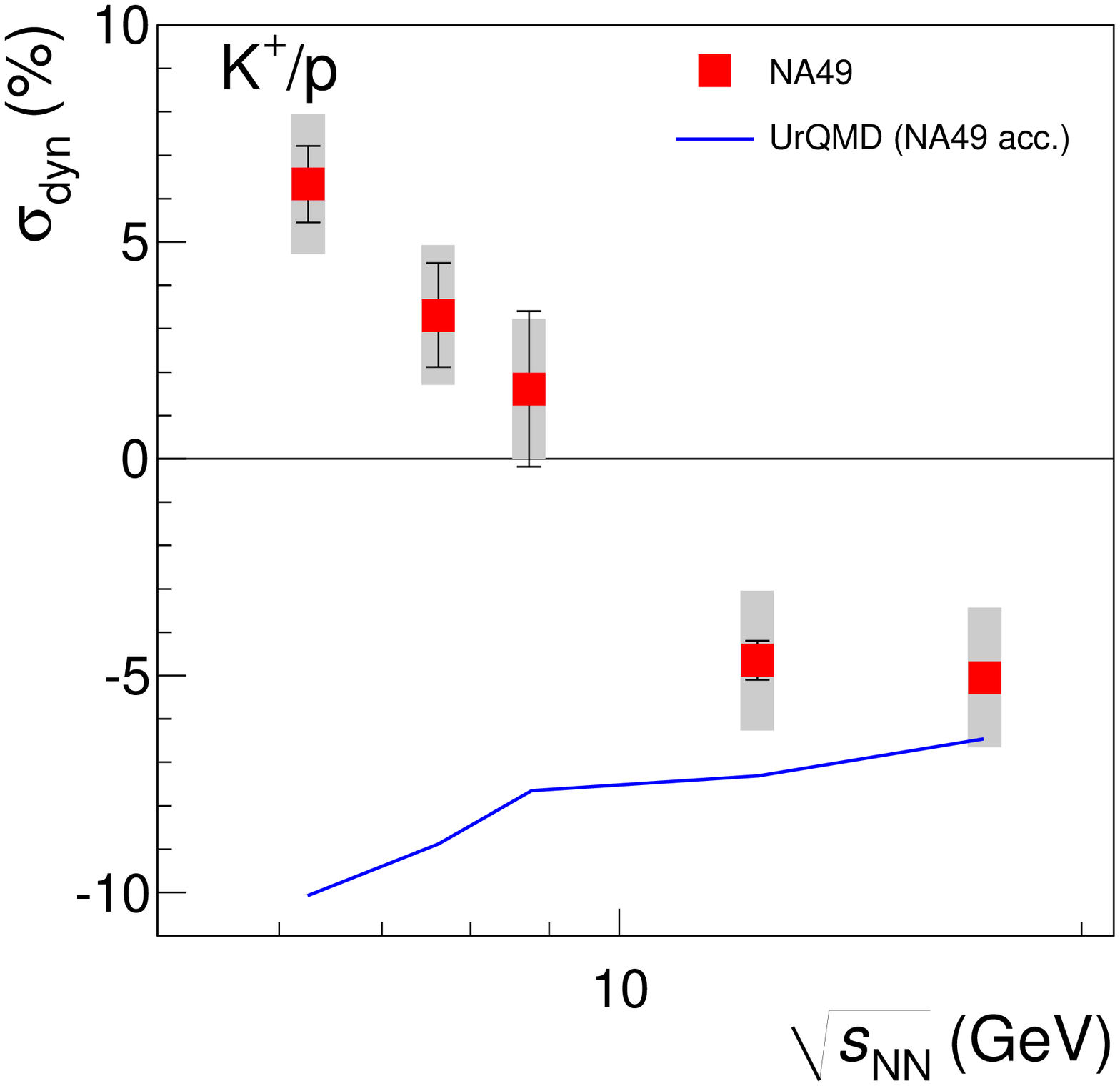}
\caption{Energy dependence of \sdyn\ for the \ktop\ and \ktopplus\ ratios in central Pb+Pb collisions, compared to calculations in the transport models UrQMD and HSD, as well as to the multiplicity scaling described in the text.}
\label{fig:Exc_KPr}
\end{figure}

In a new analysis~\cite{Anticic:2011am} fluctuations of the kaon to proton ratio were investigated, motivated by their conjectured connection to the baryon-strangeness correlation~\cite{Koch:2005vg}.
The results are shown in figure~\ref{fig:Exc_KPr}. In addition to the combined charges ratio (\ktop), \sdyn\ was also evaluated for the \ktopplus\ ratio. The latter has the advantage that one source of correlation can be excluded in this channel: No resonance feeds into $\mathrm{K}^+ + \mathrm{p}$.

For both ratios, \sdyn\ shows a strong dependence on \roots, going from positive values at low energies to $\sdyn < 0$ at high energies. This behaviour is not reproduced by the transport models. HSD and UrQMD disagree with regard to the predicted value of \sdyn, and both models show practically no energy dependence in strong contrast to the data.

The scaling that successfully described the centrality dependence of all three ratios studied above as well as the energy dependence of \ktopi\ and \ptopi\ fluctuations fails to work here. The change of sign observed for \ktop\ and \ktopplus\ excludes any simple scaling based on average multiplicites.
Keeping in mind that such a scaling assumes an invariant correlation strength, this result indicates that the underlying correlation between kaons and protons is changing with energy.

\section*{References}

\end{document}